\documentclass[letterpaper,twocolumn]{jpsj3}
\usepackage{txfonts}
\usepackage{braket}
\usepackage{amsmath}
\usepackage{txfonts}
\usepackage{bm}
\usepackage{here}
\usepackage{caption}

\title{Superconducting Properties of Heavy Fermion UTe$_2$ Revealed by $^{125}$Te-nuclear Magnetic Resonance}

\author{Genki Nakamine$^1$\thanks{nakamine.genki.88v@st.kyoto-u.ac.jp}, Shunsaku Kitagawa$^1$, Kenji Ishida$^1$\thanks{kishida@scphys.kyoto-u.ac.jp}, Yo Tokunaga$^2$, Hironori Sakai$^2$, Shinsaku Kambe$^2$, Ai Nakamura$^3$, Yusei Shimizu$^3$, Yoshiya Homma $^3$, Dexin Li$^3$, Fuminori Honda$^3$, and Dai Aoki$^{3,4}$}
\inst{$^1$Depertment of Phisics, Kyoto University, Kyoto, 606-8502, Japan \\
$^2$ASRC, Japan Atomic Energy Agency, Tokai-mura, Ibaraki 319-1195, Japan \\
$^3$IMR, Tohoku University, Oarai, Ibaraki, 311-1313, Japan \\
$^4$University Grenoble, CEA, IRIG-PHERIQS, F-38000 Grenoble, France} 

\abst{We have performed the $^{125}$Te-nuclear magnetic resonance (NMR) measurement in the field along the $b$ axis on the newly discovered superconductor UTe$_2$, which is a candidate of a spin-triplet superconductor. 
The nuclear spin-lattice relaxation rate divided by temperature $1/T_1T$ abruptly decreases below a superconducting (SC) transition temperature $T_c$ without showing a coherence peak, indicative of UTe$_2$ being an unconventional superconductor. 
It was found that the temperature dependence of $1/T_1T$ in the SC state cannot be understood by a single SC gap behavior but can be explained by a two SC gap model.  
The Knight shift, proportional to the spin susceptibility, decreases below $T_c$, but the magnitude of the decrease is much smaller than the decrease expected in the spin-singlet pairing.
Rather, the small Knight-shift decrease as well as the absence of the Pauli-depairing effect can be interpreted by the spin triplet scenario.	}   

\begin{document}
\maketitle

Superconducting (SC) phase of uranium-based compounds UGe$_2$\cite{Saxena2000}, URhGe\cite{Aoki2001} and UCoGe\cite{PhysRevLett.99.067006} overlaps or is inside the ferromagnetic (FM) phase and their upper critical field $H_{\rm c2}$ is far beyond the ordinary Pauli paramagnetic limit. 
In addition, it was reported that their superconductivity is strongly coupled with the FM spin fluctuations (SFs) which are considered as a pairing interaction of the SC pairs\cite{PhysRevLett.108.066403, PhysRevLett.114.216401}.
These experimental results strongly suggest that spin-triplet superconductivity is realized in these compounds\cite{AokiJPSJRev2019,MineevRev2017}

Recently, Ran {\it et al.} discovered a new U-based superconductor UTe$_2$ with a SC transition temperature $T_c \sim$1.6 K\cite{RanScience2019}.
Although the superconductivity occurs in the paramagnetic (PM) state, SC properties in UTe$_2$ is similar to those in the above FM superconductors; 
$H_\text{c2}$ in UTe$_2$ is extremely large and anisotropic, and exceeds the Pauli-limiting field along all three principal axes\cite{RanScience2019, doi:10.7566/JPSJ.88.043702}.
More interestingly, superconductivity in UTe$_2$ becomes enhanced and $T_c$ increases when magnetic field ($H$) along the $b$ axis (the magnetic hard axis) is greater than 15 T\cite{RanScience2019, KnebelJPSJ2019}.  
Recent $^{125}$Te NMR measurement on a single-crystalline UTe$_2$ revealed that SFs above 20 K is FM and shows moderate Ising anisotropy\cite{doi:10.7566/JPSJ.88.073701}. 
Therefore, UTe$_2$ is also considered as a promising candidate of a spin-triplet superconductor realized in the PM state. 
Another intriguing feature in UTe$_2$ is the large residual Sommerfeld coefficient term in the SC state, which is approximately half of the normal-state value at $T_c$\cite{RanScience2019, doi:10.7566/JPSJ.88.043702}.
This is interpreted by the fascinating scenario of a non-unitary triplet state, in which only a specific spin (up-up or down-down spin) on the Fermi surfaces is superconducting. 
However, time reversal symmetry breaking which originates from spontaneous magnetization has not been found from zero-field muon relaxation rate in the SC state,\cite{2019arXiv190506901S} and the recent specific heat measurement suggests point node gap accompanied with a quantum critical fluctuation.\cite{2019arXiv190801069M}
In addition, it is theoretically pointed out that the direct transition from the PM state to the non-unitary triplet state is unlikely\cite{Mineev}.  

Although an unconventional SC state is expected in UTe$_2$ as described above, few experimental results clarifying the SC symmetry have been reported yet\cite{arXiv:1908.01069}.  
We have performed nuclear magnetic resonance (NMR) measurement to investigate the SC properties of UTe$_2$, since NMR is a powerful probe for this purpose.

Here, we report the results of $1/T_1T$ and NMR Knight shift measurements under $H$ along the $b$ axis. 
The $1/T_1T$ result in SC state indicates that UTe$_2$ is an unconventional superconductor with a multi-gap character. 
The Knight shift decreased in the SC state, but the magnitude of the decrease was much smaller than the decrease expected in a spin-singlet pairing. 
In addition, anomalous broadening of the NMR spectrum was observed at low temperatures. 
We consider that these results as well as the large $H_{\rm c2}$ would be interpreted with a specific spin-triplet scenario. 
\begin{figure}
\includegraphics[width = 80mm]{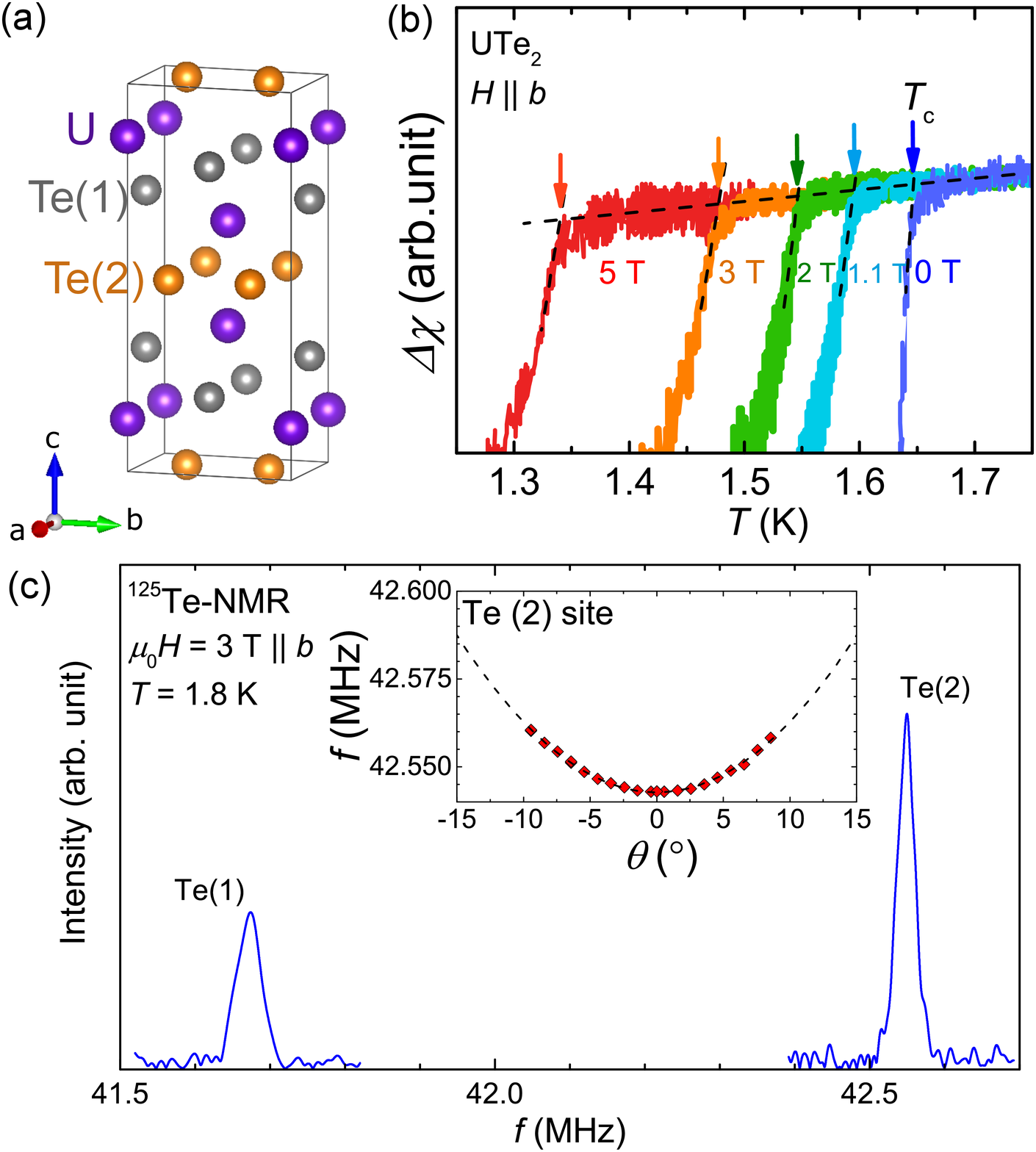}
\caption{(Color online) (a) Schematic view of the crystal structure of UTe$_2$. (b) Temperature dependence of the AC susceptibility $\Delta \chi$ at several magnetic fields along the $b$ axis. The measurement was done with an NMR tank circuit on cooling. 
(c) The typical $^{125}$NMR spectra measured in the field along the $b$-axis.   (inset) The angular dependence of the resonance frequency of the NMR signal at the Te(2) site. The dash line represents the fitting with the theoretical calculation.}
\label{f1}
\end{figure}

$^{125}$Te-NMR measurements were performed on a single crystal of $4\times2\times1$ mm$^3$ size.
Details of the  experimental set-up are described in the supplemental materials.\cite{Supplemental}
Figure \ref{f1} (a) shows the crystal structure of UTe{$_2$}.
There are two crystallographically inequivalent Te sites: 4$j$, and 4$h$ sites with point symmetries $mm2$, and $m2m$, respectively. 
We denote these sites as Te(1) and Te(2) as in the previous paper\cite{doi:10.7566/JPSJ.88.073701} respectively, which are  shown in Fig. 1(a). 
Since the peak frequency of the NMR spectrum at low temperatures depended on the energy of the RF-pulses due to the Joule-heating, the NMR spectrum was obtained by the small-energy RF pulses enough not to show the energy dependence of the peak frequency.
The detailed dependence of the NMR-peak frequency on the energy of the RF pulses is shown in the supplemental material.\cite{Supplemental}

\begin{figure}
\includegraphics[width = 80mm]{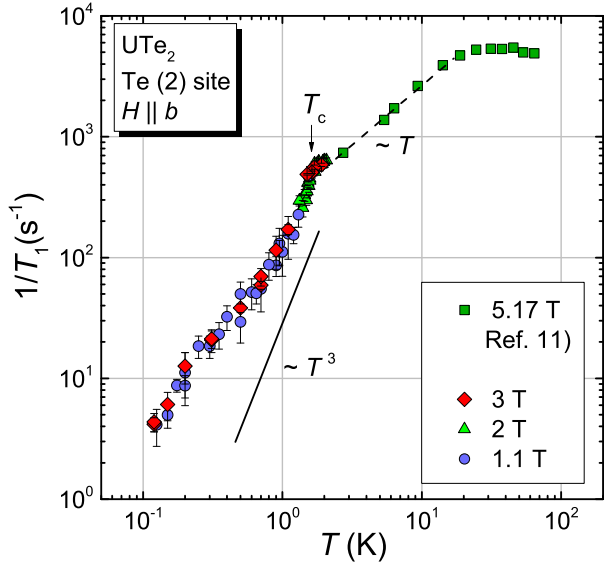}
\caption{The temperature dependence of $1/T_1$ at the Te(2) site measured in $H$ along the $b$ axis. The $1/T_1$ below 2 K was obtained from this experiment.
The squares represent $1 / T_1$ in the normal state reported by Tokunaga $et$ $al$\cite{doi:10.7566/JPSJ.88.073701}. }
\label{f2}
\end{figure}

\begin{figure}
\includegraphics[width = 80mm]{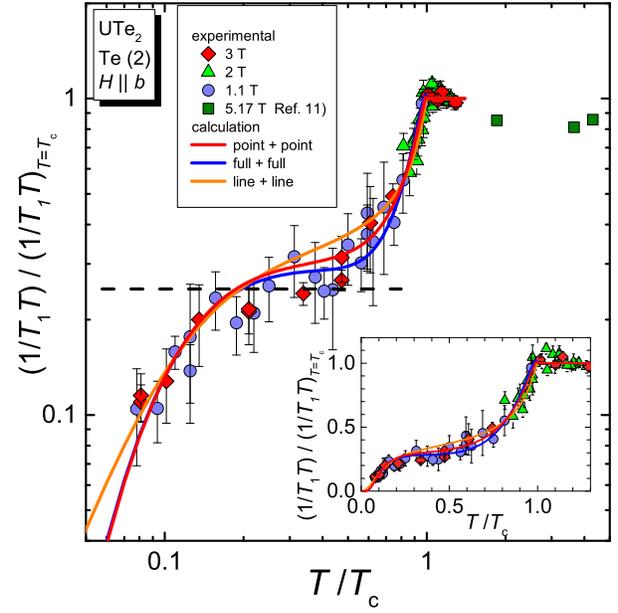}
\caption{(Color online) Plot of $1/T_1T$ normalized by the value at $T_c$ against $T/T_c$ in the log-log scale. The inset shows the linear scale.
Calculations of normalized $1/T_1T$ by the two SC-gap model with three different SC gap structures are shown with curves.}
\label{f3}
\end{figure}

The AC susceptibility measurements by recording the tuning of the NMR coil with the sample were also carried out and $T_c$ was determined as shown in Fig. \ref{f1}(b). 
Figure \ref{f1}(c) shows the $^{125}$Te-NMR spectra of both the Te(1) and Te(2) sites measured at 1.8 K under 3 T along the $b$ axis.  
It was reported that the spectrum at higher [lower] frequency was assigned as the Te(2) [Te(1)] site\cite{doi:10.7566/JPSJ.88.073701}. 
The inset of Fig.~1 (b) showed the angular dependence of the resonance peak of the Te(2) NMR spectrum in the $bc$ plane. 
The field was aligned to the $b$ axis (defined as 0 $^\circ$), and its misalignment is less than 0.5 $^\circ$.
The temperature ($T$) dependence of $1/T_1$ measured at 1.1, 2 and 3 T in $H \parallel b$ is shown in Fig. \ref{f2}, in which the normal-state $1/T_1$ measured at 5.17 T is also plotted\cite{doi:10.7566/JPSJ.88.073701}. 
As reported previously\cite{doi:10.7566/JPSJ.88.073701}, $1/T_1$ was almost constant down to 20 K, below which the resistivity and $1/T_1$ show a metallic behavior. 
These are typical heavy-fermion behaviors, observed also in UPt$_3$\cite{KohoriJPSJ1988} and UCoGe\cite{OhtaJPSJ2010}. 
Below 10 K, $1/T_1$ was proportional to $T$ and suddenly decreased at $T_c$ due to the occurrence of superconductivity. 
One of the remarks is that the coherence peak just below $T_c$, which is a hallmark of a conventional $s$-wave superconductivity, was not observed.
This shows that UTe$_2$ is an unconventional superconductor. 
Another important feature is that $T$ dependence of $1/T_1$ could not be expressed with a single SC gap but showed a multi-SC gap behavior.
This is recognized from Fig. \ref{f3}, in which $T$ dependence of $1/T_1T$ normalized by the values at $T_c$ are plotted in a log-log scale.  
$1/T_1T$ suddenly decreased in the SC state and saturated at around the value of the dashed line, which corresponds to a quarter of the normal-state value. 
Taking into account that $1/T_1T$ is proportional to the square of the density of states (DOS), this behavior indicates that the DOS at 0.2 K is a half of the normal state value. 
The $T$ dependence of $1/T_1T$ down to 0.2 K seems to be consistent with that of the specific-heat result.  
It seems that an additional decrease was observed below 0.2 K, suggestive of the presence of another small SC gap.
However, the small-gap behavior was not changed by $H$ within the measurement range. 
In order to understand above-mentioned anomalous $T$ dependence of $1/T_1$ quantitatively, we analyzed the experimental data by assuming a two-SC-gap model with three kinds of the gap, i.e. point or line node gap or full gap\cite{KitagawaPRB2017}. 
In this model, the $T$ dependence of $1/T_1$ is expressed as
\begin{equation*}
\begin{split}
1/T_1 \propto &\int_0^\infty\left[\left(\sum_{i=1,2}n_i N^i_s(E)\right)^2+\left(\sum_{i=1,2}n_i M^i_s(E)\right)^2\right] \\
&\times f(E)\left[1-f(E)\right]dE
\end{split}
\end{equation*}
where $N^i_s(E)$, $M^i_s(E)$, and $f(E)$ are the quasiparticle DOS, anomalous DOS originating from the coherence effect of the SC pairs, and the Fermi distribution function, respectively. 
The $n_i$ represents the fraction of the DOS of the $i$-th gap and $n_1+n_2 =1$ holds. 
Since the value of $1/T_1T$ at 0.2 K is close to a quarter of the normal state value, we assumed that $n_1 = n_2 = 0.5$. 
Furthermore, we take $M^i_s = 0$ because of the absence of the coherence peak. 
Details of the calculation are described in the supplemental materials. \cite{Supplemental}

\begin{table}[htb]
  \begin{center}
    \caption{Magnitude of superconducting gaps $\Delta_i$ divided by $k_BT_c$, and smearing factor $\delta$ used for the calculation of $T_1$ by the two SC-gap model with three different SC gap structures.}
    \begin{tabular*}{80mm}{@{\extracolsep{\fill}}lccc} \hline \hline
      Model & $\Delta_1/k_BT_c$ & $\Delta_2/k_BT_c$ & $\delta$ \\ \hline \hline
      full gap & 3.5 & 0.24 & 0.02 \\
      point node & 3.8 & 0.28 & - \\
      line node & 5.9 & 0.36 & - \\ \hline
    \end{tabular*}
    \label{tab1}
  \end{center}
\end{table}
The calculated $1/T_1T$ with each gap is shown in Fig. \ref{f3}.
The experimental data of $1/T_1T$ can be reproduced by the model with the parameters listed in Table I. Here, $k_B$ is Boltzmann's constant.
Since $\Delta_1/k_\text{B}T_c$ exceeds the BCS value of 1.76 in all gaps, it is considered that the superconductivity is in the strong-coupling regime, which is also suggested from a large jump at $T_c$ in the specific-heat measurement. 
However, the parameter of $\Delta_1/k_\text{B}T_c$ of the line node model for reproducing a sharp decrease of $1/T_1T$ just below $T_c$ is too large (5.9), and thus the line node seems to be less likely. 
This might be consistent with the spin-triplet scenario, since it was pointed out from the theoretical study that the line nodal odd-parity superconductivity is unstable in a symmorphic system\cite{BlountPRB1985}.
In the all SC gap models, the reduction of $1/T_1T$ below 0.2 K could be reproduced by the presence of the smaller gap $\Delta_2$, the size of which is approximately one tenth of $\Delta_1$. 
Although remarkable two SC gap behavior has not been observed in other measurements at present\cite{arXiv:1908.01069}, the multi-SC gap character seems consistent with the recent first-principle calculation and ARPES result showing the presence of electron and hole Fermi surfaces\cite{IshizukaArXiv, FujimoriJPSJ}. 

\begin{figure}
\includegraphics[width = 85mm]{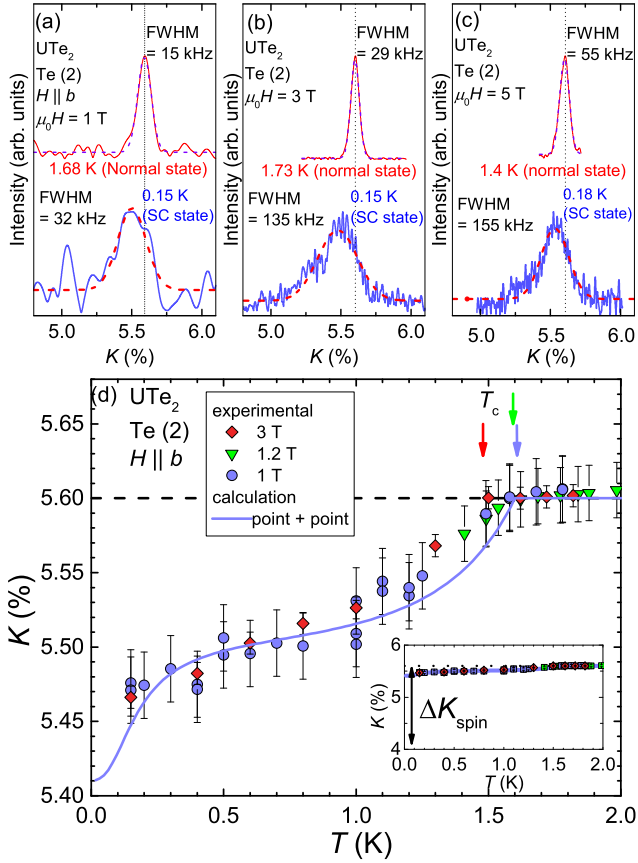}
\caption{(Color online) Typical ${\rm ^{125}}$Te-NMR spectra at the Te (2) site in the normal and SC states, measured under $H$ of (a) 1 T,  (b) 3 T and (c) 5 T along the $b$-axis, respectively. 
The dashed curves are the fitting of the NMR spectra by a Gaussian function.
(d)Temperature dependence of Knight shift measured at several magnetic fields along the $b$ axis. 
The dashed line denotes Knight shift at $T_c$. 
The solid curve is the calculation of Knight shift with a two-SC gap model of two point-node gaps with parameters listed in Table  I. 
The inset is temperature dependence of Knight shift compared with $\Delta K_{\rm spin}$ estimated from the eq (A).  }
\label{f4}
\end{figure}
Figures \ref{f4} (a),  (b) and (c) are  typical ${\rm ^{125}}$Te-NMR spectra on the Te(2) site in the normal and SC states in 1, 3 and 5 T along the $b$-axis, respectively.
The NMR spectrum shifts to a lower frequency side and is broadened in the SC state.
It is well-known that the linewidth of the NMR spectrum increases in the SC state due to the SC diamagnetic effect, but this effect was estimated to be $\Delta f \sim 1.2$ kHz from the $\mu$SR measurement\cite{2019arXiv190506901S}, which cannot explain the broadening we observed.  
It seems that the linewidth broadening at 0.15 K is not due to the conventional SC diamagnetic effect, but rather intrinsic properties of its superconductivity. 
  
Figure 4 (d) shows the temperature dependence of the Knight shift evaluated from the fitting of the NMR spectrum measured in 1, 1.2 and 3 T along the $b$-axis. 
It was found that Knight shift decreases below $T_c$ and the decreasing rate once saturates in the temperature region between 1 and 0.3 K as seen in $1/T_1T$ in Fig.~3. 
In general, the decrease of the Knight shift in the SC state is represented as $\Delta K = \Delta K_{\rm spin} + \Delta K_{\rm dia}$, where $\Delta K_{\rm spin}$ is the decrease in a spin part of the Knight shift, and $\Delta K_{\rm dia}$ is the decrease by the SC diamagnetic effect.
As already discussed, it is considered that the magnitude of $\Delta K_{\rm dia}$ is small, since $\Delta K_{\rm dia}$ should be smaller in the larger field but no appreciable difference in $\Delta K$ was observed between 1 and 3 T. 
In fact, $\Delta K_{\rm dia}$ was estimated as $\sim$ 0.035\% at 1 T from the $\mu$SR experiment\cite{2019arXiv190506901S}, and thus would be negligibly small.
Therefore, in the later discussion, we assume that $\Delta K$ arises from $\Delta K_{\rm spin}$.
The magnitude of $\Delta K$ was roughly estimated to be $\sim$ 0.13\% from the $T$ extrapolation. 
More quantitatively, $\Delta K$ can be estimated from the fitting of the calculated $K$ with the two-SC-gap models used in the $1/T_1T$ fitting, since Knight-shift in the SC state is also related with the quasiparticle DOS as in the case of $1/T_1T$, and is expressed as
\[
\frac{K}{K_{\rm n, spin}} \propto \sum_{i = 1,2}n_i~ \int_{0}^{\infty}N_s^i~\frac{\partial f(E)}{\partial E}dE,
\]
where $K_\text{n, spin}$ is the spin-part of the Knight shift, which is proportional to the spin-susceptibility related with the SC pair.  
It is noted that $K$ should decrease to zero in a spin-singlet pairing.
The calculated $K$ seems to reproduce the experimental result of $K$ consistently, and $\Delta K$ is estimated to be $\sim$ 0.2\% from the comparison to the whole behavior of the experimental results shown in Fig. \ref{f4} (d).
     
Now we discuss the relationship between the normal-state $K$ and the decrease of $K$ in the SC state.
In NMR studies of transition metals, the spin-part of the Knight shift has been evaluated from the $K-\chi$ plot.\cite{ClogstonPhysRev1964}
However, it was clarified that this simple estimation is not valid in heavy-fermion superconductors due to the strong spin-orbit coupling, particularly for the estimation of the spin-part $K$ related to the superconductivity.  
In such a case, $\Delta K$ related with the superconductivity can be estimated from the change of the Sommerfeld coefficient $\gamma_{\rm el}$ in the SC state as\cite{doi:10.1143/JPSJ.74.1245} 
\[
\Delta~K_{\rm spin} = \frac{A_{\rm hf}}{N_A \mu_B}\Delta\chi^{\rm qp}=\frac{A_{\rm hf}}{N_A \mu_B}\frac{\Delta\gamma_{\rm el}(\mu_{\rm eff})^2}{\pi^2k_B^2}R.
\]
Here, $A_{hf}$ is a hyperfine coupling constant, $N_A$ is Avogadro's number, $\mu_B$ is Bohr magneton, $\Delta\chi^{\rm qp}$ is the change of the quasi-particle spin susceptibility below $T_c$, $\mu_{\rm eff}$ is the effective moment and $R$ is the Wilson ratio. 
Actually, $\Delta K_{\rm spin}$ determined by this method was in good agreement with the experimental results of $\Delta K$ observed in UPd$_2$Al$_3$\cite{doi:10.1143/JPSJ.74.1245} and URu$_2$Si$_2$\cite{PhysRevLett.120.027001}, which are spin-singlet superconductors. 
In UTe$_2$, if we assume that the whole part of the $\gamma_\text{el}$ diminishes at $T = 0$, $\Delta K_{\rm spin}$ is estimated as $\sim 1.49 \times R$\%, where the experimental value of $\gamma_{el} \sim$ 117 mJ/mol$\cdot$K$^{2}$\cite{doi:10.7566/JPSJ.88.043702} and $A_{\rm hf} \sim$ 51.8 kOe / $\mu_B$\cite{doi:10.7566/JPSJ.88.073701} determined from a $K-\chi$ plot are used.
Although the Wilson ratio in UTe$_2$ is difficult to estimate due to the large anisotropy of the spin susceptibility at low temperatures, it is known to take a value of 1 $\sim$ 2 in heavy-fermion compounds.
The experimental reduction $\Delta K$ is approximately one order of magnitude smaller than the calculated $\Delta K$, suggestive of the large residual spin component at $T$ = 0. 
Since the magnetic fields of the Knight-shift measurements (1 and 3 T) are much smaller than $H_\text{c2}$, the presence of such a large spin component seems incompatible with the spin-singlet, but supports the triplet scenario. 
We point out that such a tiny decrease of the Knight shift in the SC state was also reported in UPt$_3$, where the decrease in the SC state measured in low fields ($\sim 0.2$ T) is only $1 \sim 2$\% of the normal-state Knight shift values\cite{TouPRL1998}. 

However, even in the spin-triplet scenario, following remarks are needed. 
Since the $b$ axis is a magnetic hard axis, the $b$ axis has the largest component of the ${\bm d}$ vector among three crystalline axes, where the spin components in the spin-triplet state are perpendicular to the ${\bm d}$ vector.  
When ${\bm d} \parallel {\bm H}$, the spin-part Knight shift largely decreases in this condition, if the ${\bm d}$ vector is strongly fixed to a crystalline axis by the spin-orbit interaction. 
However, the observed $\Delta K$ is too small to explain the estimated spin component. 
Therefore, the multicomponent of the ${\bm d}$ vectors and/or the ${\bm d}$ vector rotation by the small field are pointed out as a possible explanation for the small $\Delta K$.  
In any case, further experiments are needed to clarify the spin state of the SC state, particularly the Knight-shift measurement along other crystalline axes or under the larger fields above 15 T is crucially important.

Finally, we comment on the broadening of the linewidth of the NMR spectrum of $^{125}$Te in the SC state.
The broadening cannot be explained by the conventional SC diamagnetic effect. 
In a spin-singlet superconductor, it is possible that linewidth becomes narrower in the SC state when the linewidth is determined by the inhomogeneity of the spin susceptibility, since the spin susceptibility decreases in the SC state.
Such behavior was observed in organic superconductors\cite{MayaffreNatPhys2014}.
On the other hand, the linewidth increases but the center of the gravity of the spectrum decreases in the SC state of UTe$_2$.
Furthermore, the broadening at 0.15 K is anisotropic and nearly proportional to $H$, suggesting that the broadening is ascribed to the large residual spin susceptibility at low temperatures.
We speculate on that the broadening might originate from the texture structure produced by the spin-triplet pairs with the spin and orbital degrees of freedom.      
It is also important to follow the field dependence of the lineshape and linewidth of the NMR spectrum up to $H_{c2}$. 

In summary, the temperature dependence of $1/T_1T$ indicates that UTe$_2$ is the unconventional superconductor with the multi-gap character. 
The Knight shift under $H$ along the $b$ axis decreases in the SC state, but the observed decrease is much smaller than the decrease expected in the spin-singlet pairing.
We consider that these results would be explained by the spin-triplet scenario with the multicomponent and/or field dependent ${\bm d}$ vectors rather than the spin-singlet pairing.
However, further experiments are needed for thorough understanding of the SC properties of UTe$_2$.    

\begin{acknowledgment}

\acknowledgments
The authors would like to thank M. Manago, T. Taniguchi, J. Ishizuka, Y. Yanase, Y. Maeno and S. Yonezawa, for valuable discussions. 
This work was supported by the Kyoto University LTM Center, Grants-in-Aid for Scientific Research (Grants No. JP15H05745, JP17K14339, JP19K03726, JP16KK0106, JP19K14657, and JP19H04696), and Grants-in-Aid for Scientific Research on Innovative Areas "J-Physics" (Grants No. JP15H05882, JP15H05884, and JP15K21732). 

\end{acknowledgment}

\bibliography{17619}

\onecolumn

\renewcommand{\figurename}{Fig. Suppl}
\setcounter{figure}{0}
\section*{Supplemental}
\section{Experimental procedure}
\begin{figure}[H]
\begin{center}
\includegraphics[width = 100mm]{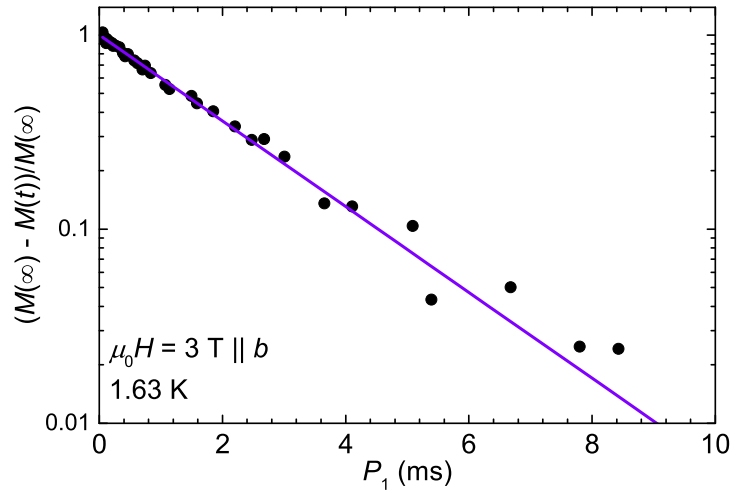}
\end{center}
\captionsetup{labelformat=empty,labelsep=none}
\caption{Typical relaxation curve $R$($t$) of the nuclear-spin magnetization after the saturation of the nuclear magnetization of $^{125}$Te and the fitting by the single exponential function at 1.63 K under 3 T along the $b$ axis.     } 
\label{fs1}
\end{figure}

Single-crystal UTe{$_2$} was grown using the chemical transport method with Iodine as a transport agent.   
$^{125}$Te-NMR measurements were performed on a single crystal of the $4\times2\times1$ mm$^3$ size. 
The $^{125}$Te nucleus (natural abundance:  7\%) has a spin $1/2$ with the gyromagnetic ratio $^{125}\gamma/2\pi = 13.454$ MHz/T.
The $^{125}$Te-NMR spectrum was obtained by the Fourier transform of a spin-echo signal observed after the spin-echo RF pulse sequence. 
We used the split superconducting (SC) magnet generating a horizontal field and a single-axis rotator with the $a$-axis being the rotation axis, in order to apply $H$ exactly parallel to the $b$ axis.
The $^{125}$Te nuclear spin-lattice relaxation rate $1/T_1$ was determined by fitting the time dependence of the spin-echo intensity after the saturation of the nuclear magnetization $M$. 
The fitting function is a simple exponential function in the case of $I$ = 1/2, and the typical relaxation curve $R(t)$ of the nuclear magnetization defined as $R(t) \equiv [M(\infty)- M(t)]/M(\infty)$ at time $t$ after the saturation of $M$ and the fitting for $T_1$ determination are shown in Fig. Suppl \ref{fs1}.   

The magnetic field was calibrated by the $^{65}$Cu ($^{65}\gamma/2\pi = 12.089$ MHz/T) NMR signal arising from the NMR coil.
Low-temperature NMR measurements down to 100 mK were performed with a ${^3}$He-${^4}$He dilution refrigerator, and the temperature was confirmed by the $T_1$ measurement of the $^{65}$Cu of the coil at 100 mK.
The AC susceptibility measurements by recording the tuning of the NMR coil were also carried out to determine $T_c$ of the sample under $H$. 

\section{Heat-up effect by the radio-frequency pulses for the NMR measurement}
In NMR measurements at low temperatures, particularly on SC compounds, the sample is heated up by the irradiation of radio frequency (RF) pulses for the NMR measurements. 
This is called ``heat-up'' effect. 
If such an effect is large, it is not possible to obtain reliable results from the NMR measurement. 
We performed several heat-up tests to check whether our measurements have been done properly or not. 

One of tests is for the $T_1$ measurement in the SC state.
Figure Suppl \ref{f2} (a) shows the pulse sequences for the $T_1$ measurement. 
The first pulse is for exciting the nuclear spin to the upper spin level, and following two pulses after time $P_1$ are for observing the spin-echo signal.
The intensity of the spin echo signal at time $t$, proportional to the nuclear magnetization $M$($t$), is related to the number of nuclear spins relaxing during $P_1$, and thus the relaxation time $T_1$ is measured by the recovery of the spin-echo intensity.
In this process, the heat-up effect by the first pulse is needed to be checked since the first pulse makes the temperature of the sample increase. 
Therefore, we examined how long is needed for the temperature heated up by the first pulse to return to the original temperature by using the same circuit for the $T_1$ measurement.
Figure Suppl  \ref{f2}  (b) shows pulse sequence for the heat-up test. 
The first pulse is the same condition with that used in the $T_1$ measurement. 
The small-energy RF pulse (called phase-detection pulse) is irradiated after $P_1$. 
If the temperature is increased by the irradiation of the first pulse, the inductance of the NMR-tank circuit is changed since the SC Meissner screening is suppressed by the heat-up effect due to temperature increase.
Therefore, we can know the heat-up effect from the intensity of the phase-detection pulse, since the intensity after the NMR receiver is sensitive of the tuning of the NMR  tank circuit. 
Figures Suppl  \ref{f2}  (c, d) show the $P_1$ dependence of the intensity of the phase-detection pulse at 1 K and 0.15 K.  
The power of the first pulse is 201 $\mu $J for 1 K and 127 $\mu $J for 0.15 K. 
Dashed lines represent the original intensity of the detection pulse without the first pulse. 
At both 1 K and 0.15 K,  the sample temperature heated up by the irradiation of the first pulse returns to the original temperature within 10 ms. 
In fact, a fast component was observed between 0 and 5 ms in the relaxation curve defined as $[M(\infty) - M(t)] / M(\infty)$ at 1.1 K [Fig. Suppl  \ref{f2} (e)], showing the presence of the heat-up effect. 
Thus, we did not use such a fast component and fit the $M$($t$) after 10 ms for the $T_1$ measurement.  
On the other hand, the fast component was not recognized in the recovery curve at 0.12 K [Fig. Suppl  \ref{f2} (f)].
This is presumably because $T_1$ become sufficiently longer than 10 ms at low temperature.

Next, we performed a heat-up check for the Knight shift measurement. 
Figure Suppl  \ref{f3} (a) shows a pulse sequence for observing spin echo. 
When two successive pulses are irradiated with an interval $\tau$, a spin echo signal appears at the $\tau \sim 50 \mu$s after the second pulse.
We found the shift of the resonance frequency of the spin-echo signal due to the instantaneous heat-up effect by the irradiations of the two pulses.
To suppress this heat-up effect, the energy of the two pulses are needed to be reduced. 
Therefore, we investigated how much power of these two pulses should be weakened to avoid the heat up effect. 
The pulse sequence for the test is shown in Figure Suppl  \ref{f3} (b). 
The phase-detection pulse is applied at the same position as the spin echo signal to investigate the sample temperature at the spin-echo position. 
Figures Suppl  \ref{f3}(c) and (d) show the RF-pulse energy dependence of the intensity of the phase detection pulse (squares) and resonance frequency (circles) at 0.8 K and 0.15 K, respectively. 
Dashed lines represent the original intensity of the phase-detection pulse without applying the two RF pulses for the spin-echo observation.
It is noted that the RF-pulse energy dependence of the resonance frequency roughly scales with that of the phase-detection pulse intensity, and is unchanged below 100 $\mu$J. 
It was found that the heat-up effect can be suppressed by reducing the RF-pulse energy below 100 $\mu$J at both temperatures. 
Therefore we performed Knight shift measurements with the two RF pulses with the smaller energy less than 100 $\mu$J.

\begin{figure}[H]
\begin{center}
\includegraphics[width = 150mm]{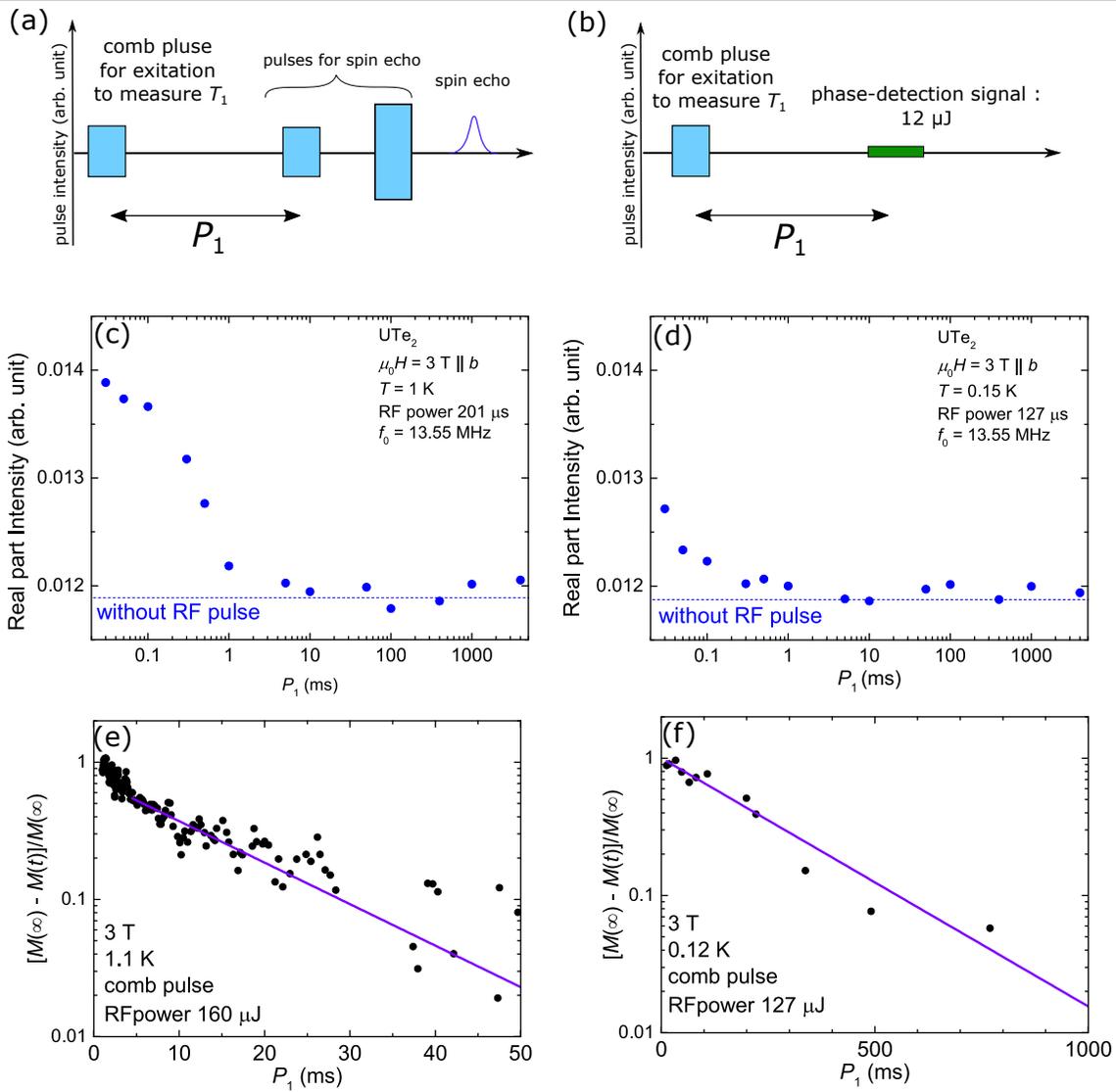}
\caption{Pulse sequence for the $T_1$ measurement (a) and the heat-up test (b). $P_1$ dependence of the intensity of the phase detection pulse after the NMR receiver at 1 K (c) and 0.15 K (d). 
The relaxation curve of the nuclear-spin magnetization, defined as $[M(\infty) - M(t)] / M(\infty)$ of $^{125}$Te nuclei at 1.1 K (e) and 0.12 K (f). }
\label{fs2}
\end{center}
\end{figure}

\newpage

\begin{figure}[H]
\begin{center}

\includegraphics[width = 150mm]{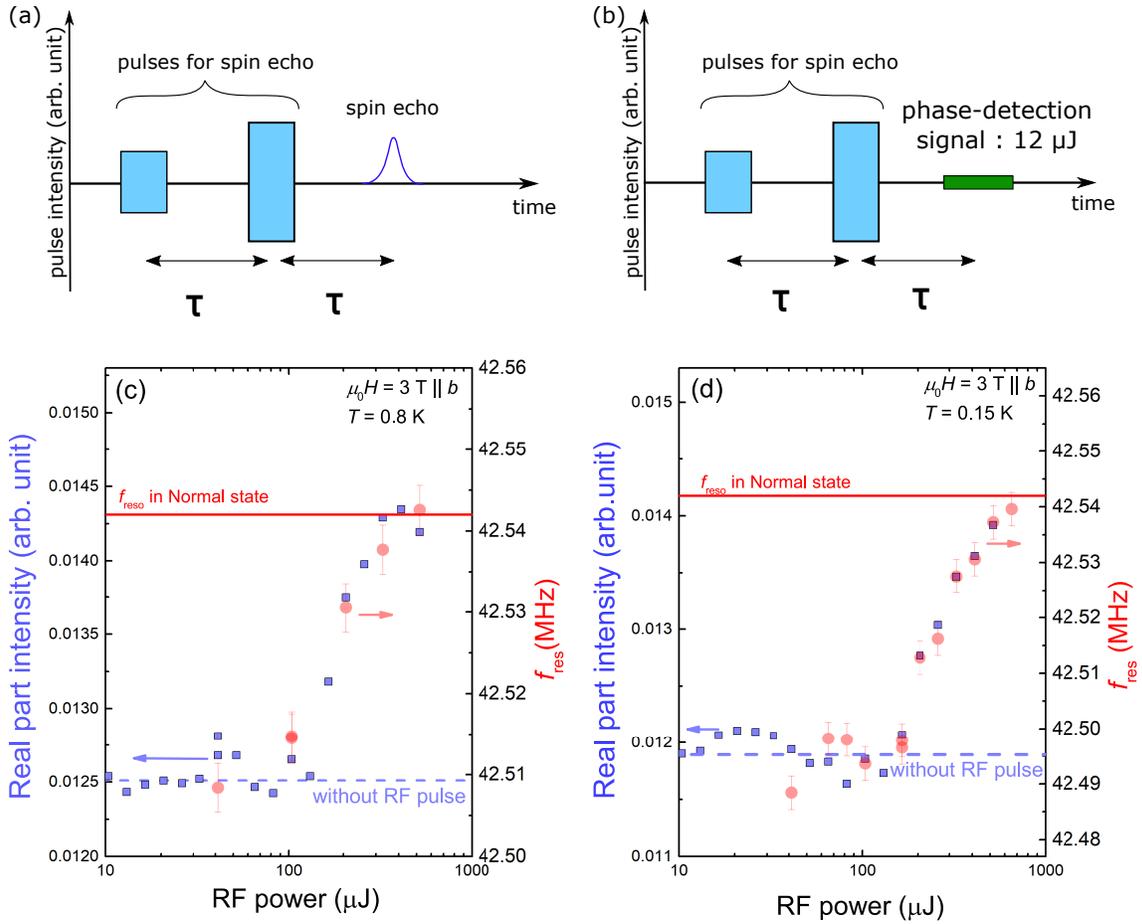}
\caption{Pulse sequence for the Knight-shift measurement (a) and the heat-up test (b). The RF-pulse energy dependence of the intensity of the phase-detection pulse after the NMR receiver (left axis) and the peak frequency of the spin-echo signal (right axis) at 0.8 K (c) and at 0.15 K (d).   The blue dash lines for the left-axis scale indicate the intensity of the phase-detection pulse without the two pulses for observing the spin-echo signal, and the red lines for the right-axis scale indicate the peak frequency of the spin-echo signal in the normal state at 1.8 K.     } 
\label{fs3}
\end{center}
\end{figure}

\section{Analyses of $1/T_1T$ by the two-SC gap model }
In order to understand anomalous $T$ dependence of $1/T_1$ in the SC state, we analyze the experimental data of $1/T_1$ by assuming a two-SC-gap model with three kinds of the gap, i.e. point- or line-node gap or full gap. 
In this model, the temperature dependence of $1/T_1$ is expressed as
\begin{equation*}
\begin{split}
1/T_1 \propto &\int_0^\infty\left[\left(\sum_{i=1,2}n_i N^i_s(E)\right)^2+\left(\sum_{i=1,2}n_i M^i_s(E)\right)^2\right]f(E)\left[1-f(E)\right]dE
\end{split}
\end{equation*}
where $N^i_s(E)$, $M^i_s(E)$, and $f(E)$ are the quasiparticle DOS, anomalous DOS originating from the coherence effect of the SC pairs, and the Fermi distribution function, respectively. 
The $n_i$ represents the fraction of the DOS of the $i$-th gap and $n_1+n_2 =1$ holds. 
Since the value of $1/T_1T$ at 0.2 K is close to a quarter of the normal state value, we assume that $n_1 = n_2 = 0.5$. 
In addition, we take $M^i_s = 0$ because of the absence of the coherence peak. 
In the calculation, the $N_s$ for respective gaps used in this calculation are expressed as following,  
\begin{eqnarray*}
{\rm full\;gap }&:& \qquad       N^i_s=Re\left(\frac{E+ig}{\sqrt{(E+ig)^2-\Delta_i^2}}\right)\\
{\rm point\;node}&:&\qquad {N^i_s}=\frac{E}{2\Delta}\ln{\left|\frac{E+\Delta_i}{E-\Delta_i}\right|}\\
{\rm line\;node}&:&\qquad   {N^i_s} = \left\{ \begin{array}{ll}
    (\frac{E} {\Delta_i})\arcsin{\frac{\Delta_i}{E}} & (E>\Delta_i) \\
   {\pi E}/{2\Delta_i} & (E<\Delta_i) .
  \end{array} \right.
\end{eqnarray*}
Here, $\Delta_i$ is the $i$-th SC gap and $g$ in the formula is a smearing factor to reduce the divergence of the DOS at $E = \Delta_i$. 

As for $T$ dependence of the SC gap, we numerically solved the formula derived from the BCS theory, which is given by 
\begin{equation*}
\frac{1}{N(0)V}=\int^{\hbar \omega_{\rm cutoff}}_0\frac{\tanh{(\sqrt{\epsilon^2+\Delta^2}/2k_BT)}}{\sqrt{\epsilon^2+\Delta^2}}d\epsilon
\end{equation*}
where $N(0)$ is DOS at the Fermi energy, and $V$ is the strength of the pairing interaction. 
Since $\Delta$ is zero at the $T_c$, $1/N(0)V$ can be calculated from the equation
\begin{equation*}
\frac{1}{N(0)V}=\int^{\hbar \omega_{\rm cutoff}}_0\frac{\tanh{(\epsilon/2k_BT_c)}}{\epsilon}d\epsilon.
\end{equation*}
The calculated temperature dependence of SC gap is shown in Fig. Suppl \ref{f4}.
The calculated $1/T_1T$ with each model is shown in Fig.~3 in the main paper, and the experimental data of $1/T_1T$ can be reproduced by the models with the parameters listed in Table I in the main paper. 
\begin{figure}[H]
\begin{center}
\includegraphics[width = 80mm]{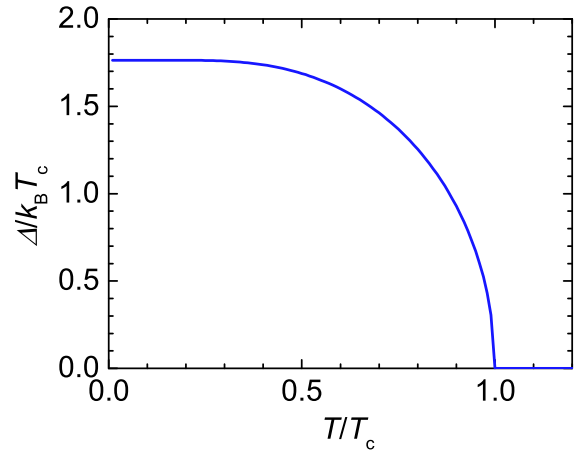}
\end{center}
\caption{The temperature dependence of the SC gap calculated by the formula derived from BCS theory.  } 
\label{fs4}
\end{figure}

\end{document}